\documentstyle[graphicx,amsmath,amsfonts,twocolumn]{article}
\bibliography{plain}
\title{Negativity as Entanglement Degree of the Jaynes-Cummings Model} \vspace{20mm}

\author{
 S. J. Akhtarshenas\thanks{E-mail: akhtarshenas@phys.ui.ac.ir}
 , M. Farsi\thanks{E-mail: mfarsi@phys.ui.ac.ir}
\\
{\small Quantum Optics Group, Department of Physics, University
of Isfahan, Isfahan, Iran } }
\begin{document}
\maketitle
%\newpage

\begin{abstract}
In this paper, by using the notion of negativity, we study the
degree of entanglement of a two-level atom interacting with a
quantized radiation field, described by the Jaynes-Cummings model
(JCM). We suppose that initially the field is in a pure state and
the atom is in a general mixed state. In this case the negativity
fully captures the entanglement of the JCM. We investigate the
case for that the initial state of the field is a coherent state.
The influences of the detuning on the degree of entanglement is
also examined.


\end{abstract}
%\pagebreak

%\vspace{7cm}

\section{Introduction}
Quantum entanglement is one of the most striking features of
quantum mechanics which has recently attracted much attention in
view of its connection with theory of quantum information and
computation. It has been recognized that entanglement provides a
fundamental potential resource for communication and information
processing \cite{ben1,ben2,ben3}. Entanglement is usually arising
from quantum correlations between separated subsystems which can
not be created by local actions on each subsystems. A pure
quantum state of two or more subsystems is said to be entangled
if it is not a product of states of each components. On the other
hand, a bipartite mixed state $\rho$ is said to be entangled if
it can not be expressed as a convex combination of pure product
states \cite{werner}, otherwise, the state is separable or
classically correlated. Peres \cite{peres} has shown that a
necessary condition for separability of a bipartite system is
that the matrix obtained by partially transposing the density
matrix $\rho$ is still positive. Horodecki et al. \cite{horo0}
have shown that this condition is sufficient for separability of
composite systems only when the dimension of the composite
Hilbert space is $2\otimes 2$ or $2\otimes 3$.

Many efforts have been devoted to quantify entanglement,
particularly for mixed states of a bipartite system, and a number
of measures have been proposed, such as entanglement of
formation, relative entropy of entanglement and negativity. For a
mixed state, the entanglement of formation (EoF) is defined as the
minimum of average entropy of the state over all pure state
decompositions of the state \cite{ben3,woot}
\begin{equation}\label{EoF1}
E_f(\rho)\equiv\min\sum_i p_i E(\psi_i),
\end{equation}
where $E(\psi_i)=-{\rm Tr}(\rho_{i}^{A}\ln{\rho_i^A})$ is the
entanglement of the pure state $|\psi_i\rangle$, and
$\rho_i^A={\rm Tr}_{B}(|\psi_i\rangle\langle\psi_i|)$.

A class of distance measures suitable for the entanglement
measures are also introduced by Vedral et al. in
\cite{ved1,ved2}, among them is the so called relative entropy of
entanglement (REE), which for a given state $\rho$ is defined by
\begin{equation}\label{REE}
E(\rho)\equiv \min_{\sigma\in {\mathcal D}} S(\rho\parallel\sigma),
\end{equation}
where
\begin{equation}\label{RE}
S(\rho\parallel\sigma)={\rm Tr}\{\rho\ln \frac{\rho}{\sigma}\},
\end{equation}
is the quantum relative entropy, and ${\mathcal D}$ is the set of
all separable states. Equation (\ref{REE}) tells us that the
amount of entanglement in $\rho$ is its (minimum) distance from the set of
separable states. It is clear that the calculation of REE as well
as EoF needs the minimization procedure which, in general, is a
difficult task to handle.

The Peres-Horodecki criterion for separability \cite{peres,horo0}
leads to a natural computable measure of entanglement, called
negativity  \cite{zycz1,zycz2,vidal}. The negativity is based on
the trace norm of the partial transpose $\rho^{T_1}$ of the
bipartite mixed state $\rho$, and measures the degree to which
$\rho^{T_1}$ fails to be positive, i.e. the absolute value of the
sum of the negative eigenvalues of $\rho^{T_1}$
\begin{equation}
{\mathcal N}(\rho)\equiv \frac{\parallel
\rho^{T_1}\parallel_1-1}{2},
\end{equation}
where $\parallel \rho^{T_1}\parallel_1$ denotes the trace norm of
$\rho^{T_1}$. Vidal and Werner \cite{vidal} proved that the
negativity ${\mathcal N}(\rho)$ is an entanglement monotone and
therefore it is a good measure of entanglement.

A lot of works have also been devoted to the preparation and
measurement of entangled states between atom and radiation field.
It has been recognized that the Jaynes-Cummings model \cite{JC}
(JCM), comprising a two-level atom interacting with a quantized
cavity mode of radiation field, plays a central role in quantum
optics. The most interesting aspects of its dynamics, which has
received much attention, is the possible existing of entanglement
between the atom and the field. Phoenix et al \cite{phoenix} have
demonstrated that the entangled atom-field state can be derived
and have given the form of this state at all times in terms of
the eigenvalues and eigenvectors of the reduced density
operators. Furuichi et al have studied the entanglement of the
JCM with the atom, initially in a mixed state and the field in a
squeezed \cite{furuichi4} and a coherent \cite{furuichi3} state.
Scheel at al \cite{scheel} have studied the entanglement
properties of the JCM in the situation for which the atom is
initially in a mixed state, whereas the field has been prepared in
an arbitrary thermal state. They have assessed the generated
entanglement quantitatively, by evaluating the negativity and
have  found that, depending on the initial joint product state,
three different regimes occur. In \cite{raja} the authors have
studied the entanglement of the JCM in both the equilibrium and
nonequilibrium time dependent ensembles. On the basis of the
negativity of the partial transpose, Khemmani et al have shown
that the thermal state of a coupled atomic two-level system and a
field mode is never separable \cite{khem}. The evolution of the
field quantum entropy and the entanglement of the atom-field in
the JCM without the rotating-wave approximation have been also
investigated in \cite{fang}.

Considerable interest has also been devoted to the entanglement
properties of the generalized JCM to include multi-photon
interaction, nonlinearity of both the field and the
intensity-dependent atom-field coupling, stark effect and
nonresonance coupling \cite{abdel1,abdel2,abdel4}. The study of
entangling two mode thermal fields through the quantum erasing
process, in which an atom is coupled with two mode fields via the
interaction governed by the two-mode two-photon JCM has been made
in \cite{li}. In \cite{abdalla} the authors have considered the
time-dependent JCM consists of two-mode interacting with an
effective two-level atom and studied the degree of entanglement.
Entanglement between two two-level atoms interacting with a
single-mode field through a two-photon process \cite{zhou} and a
multi-photon process \cite{abdel3}  have been studied, and on the
other hand,  Wang et al have investigated the entangle dynamics
and entanglement distribution in a two cavity mode coupled to a
two-level atom via two-photon process \cite{wang}. A three-level
atom interacting with a single cavity field with an arbitrary
form of the intensity-dependent coupling has been studied in
\cite{obada}. The effect of dissipation on the entanglement
properties in the framework of the JCM and related models have
been studied \cite{zhou2,zhou3,zidan,ren}.
 A study has been done on the  entropy correlations
between an atom and a single quantized cavity mode in the
framework of the JCM by considering the both pure and mixed
atomic and field states \cite{bouk}.

There exist also a considerable interest to the problem of
ordering the density operators with respect to the amount of
entanglement \cite{eisert}.   Wei et al \cite{wei} have
determined families of maximally entangled states, i.e. the
states which posses the maximum amount of entanglement for a
given degree of mixedness. They have considered various measures
of entanglement (entanglement of formation, relative entropy and
negativity) and mixedness (linear entropy and von Neumann
entropy) and found that the form of the maximally entangled mixed
states depends on the measures used. Miranowicz et al
\cite{miran1} have studied the ordering of two-qubit states with
respect to the Wootters's concurrence \cite{woot} and the
negativity and found that the two entanglement measures can
impose different ordering on the states.

In most of the previous studies on the entanglement of the JCM,
quantum mutual entropy is adapted to measure the degree of
entanglement of the atom and the field (for instance see
\cite{furuichi4,furuichi3,abdel1,abdel2,obada,zidan}). The quantum
mutual entropy (also called index of correlation
\cite{barnett,lindblad}) for a given density matrix $\rho$ is
defined by
$$
I(\rho)\equiv\rm{Tr}\{\rho\left(\ln \rho -\ln\left(\rho^{A}\otimes
\rho^{F}\right)\right)\}
$$
\begin{equation}\hspace{1mm}
=S(\rho^A)+S(\rho^F)-S(\rho),
\end{equation}
where $\rho^{A}$ and $\rho^F$ are atomic and field density
matrices, respectively, and $S(\sigma)=-\rm{Tr}\sigma\ln{\sigma}$
is the von Neumann entropy of a given state $\sigma$. Comparing
with equation (\ref{RE}), it follows that $I(\rho)$ represents a
distance between $\rho$ and the product of marginals
$\rho^{A}\otimes \rho^{F}$, i.e. it measures all correlations
including classical as well as quantum correlation and does not
discriminate the purely quantum entanglement from the classical
correlation  \cite{ved1,ved2,cerf,ved3}.  It is evident that the
mutual entropy is bounded $0\le I(\rho)\le
2\min{[S(\rho^A),S(\rho^F)]}$, and a bipartite density matrix
with excessive mutual entropy, i.e.
$\min{[S(\rho^A),S(\rho^F)]}\le I(\rho)\le
2\min{[S(\rho^A),S(\rho^F)]}$ , is entangled. This occurs
precisely for quantum entangled systems and is forbidden for
classical systems, therefore, a necessary condition for
separability is $0\le I(\rho)\le \min{[S(\rho^A),S(\rho^F)]}$
\cite{bouk,cerf}. This means that $I(\rho)$ does not fulfill the
first property that we need from an entanglement measure; that
is, for any separable state $\sigma$ the measure of entanglement
should be zero.

In this contribution, we shall turn our attention to concentrate
on the negativity as the entanglement degree of the JCM. Since,
in practice, it is difficult to realize an atom in a pure state,
therefore we suppose that  the atom is prepared, initially, in a
general mixed state but, however, the field is in a pure state.
It is shown that in this case the whole density matrix has rank
two and supported at most on ${\mathbb C}^2\otimes{\mathbb C}^4$
space. Since all positive partial transpose (PPT) rank two
bipartite density matrices are separable \cite{horo1,kraus},
therefore the negativity fully captures the entanglement
properties of the JCM, i.e. ${\mathcal N}(\rho)=0$ iff $\rho$ is
separable. On the other hand  since the mutual entropy can not
discriminate the purely quantum entanglement from the classical
correlation \cite{cerf,ved3},  therefore there is no guarantee
that the value of the mutual entropy be precisely quantum
entanglement, and even, it may happens that for a separable state
$\rho$ one finds  a nonvanishing  mutual entropy. Indeed, when two
subsystems $A$ and $F$ are classically maximally correlated, the
classical upper bound $I(\rho)=\min{[S(\rho^A),S(\rho^F)]}$ is
saturated \cite{cerf}, although the system is quantum
mechanically disentangled. This implies that the negativity gives
a more correct estimation of the quantum entanglement. We
investigate the negativity in the case that the field is in a
coherent state $|\alpha\rangle$, and compare it with the mutual
entropy. It is shown that only when the atom is initially in a
pure state, there exist, up to a difference between the
amplitudes of oscillation, a complete agreement between the
negativity and the quantum mutual entropy, which the latter is
twice the reduced von Neumann entropy. The influences of the
detuning on the degree of entanglement is also examined.

The organization of the paper is as follows. We start by
reviewing the JCM and its solutions in section 2. The time
evolution of the system is considered in section 3. Sections 4 and
5 are devoted to the calculation of the negativity in which the
numerical results and their discussions are presented. The paper
is concluded in section 6 with a brief conclusion.

\section{The Jaynes-Cummings model}
The Hamiltonian of a two-level atom interacting with a
single-mode quantized  radiation field is described by the JCM
\cite{JC} which, within the rotating wave approximation, is one of
the few exactly solvable models in quantum mechanics. Its
dynamics exhibits such challenging features as collapse and
revival of Rabi oscillations, and also the entanglement between
the atom and the field.

The JC Hamiltonian between a two-level atom $A$ and a single-mode
quantized radiation field $F$ is described by
\begin{equation}\label{JCH}
H=H_0+H_1,
\end{equation}
where $H_0$ and $H_1$, act on the product Hilbert space ${\mathcal
H}^A\otimes{\mathcal H}^F$, are given by
\begin{eqnarray}\label{JCH0}
H_0=\frac{1}{2}\hbar \omega _A \sigma _z +\hbar \omega _F a^\dag a,
\\   \label{JCH1}
H_1  = \hbar g(\sigma_{+} \otimes a   + \sigma_{-}   \otimes
a^\dag ),
\end{eqnarray}
where $g$ is the atom-field coupling constant,
$\omega_A=(\epsilon_e-\epsilon_g)/\hbar$ is the atomic transition
frequency, and $\omega_F$ denotes the field frequency. The atomic
``spin-flip'' operators $\sigma_{+}=\left(\begin{array}{cc}0 & 1 \\
0 & 0
\end{array}\right)$, $\sigma_{-}={(\sigma_{+})}^\dag$, and the atomic inversion operator
$\sigma_z=\left(\begin{array}{cc}1& 0 \\ 0 & -1
\end{array}\right)$ act on the atom Hilbert space ${\mathcal
H}^A={\mathbb C}^2$ spanned by the ground state
$|g\rangle\rightarrow(0,1)^T$ and the excited state
$|e\rangle\rightarrow(1,0)^T$. The field annihilation and creation
operators $a$ and $a^\dag$ satisfy the commutation  relation
$[a,a^{\dag}]=1$ and act on the field Hilbert space ${\mathcal
H}^F$ spanned by the photon-number states
$\{|n\rangle=\frac{(a^\dag)^n}{\sqrt{n!}}|0\rangle\}_{n=0}^{\infty}$,
where $|0\rangle$ is the vacuum state of the field.

It is evident that the JC Hamiltonian (\ref{JCH}) conserves the
total number of excitations $K=(a^\dag a+\sigma_z/2)$. This
provides a decomposition for the system Hilbert space as
${\mathcal H}=\sum_{n=0}^{\infty}\oplus{\mathcal H}_{n}$ such that
${\mathcal H}_{0}=\{|g,0\rangle\}$ is a one-dimensional
eigensubspace of $K$ with the eigenvalue $-1/2$, and ${\mathcal
H}_{n+1}|_{n=0}^{\infty}=\{|e,n\rangle,\;|g,n+1\rangle\}$ are the
two-dimensional eigensubspaces of $K$ with the eigenvalues
$n+1/2$. Corresponding to the eigensubspace   ${\mathcal
H}_{n+1}|_{n=0}^{\infty}$, there exist the eigenvalues
$E_{\pm}^{(n)}$ and the eigenvectors $|\Phi _{\pm} ^{(n)} \rangle$
of $H$ as follows
\begin{equation}\label{Evalues}
E_{\pm}^{(n)}=\hbar \omega _F \left(n+1/2\right)\pm
\hbar\Omega_n,
\end{equation}
and
\begin{eqnarray}\label{Evectors1}
|\Phi_{+}^{(n)}\rangle=\sin{\vartheta_n}|e,n\rangle
+\cos{\vartheta_n} |g,n+1\rangle,
\\   \label{Evectors2}
|\Phi_{-}^{(n)}\rangle=\cos{\vartheta_n}|e,n\rangle -
\sin{\vartheta_n} |g,n+1\rangle,
\end{eqnarray}
with
\begin{equation}
\tan{\vartheta_n}=\frac{2 g\sqrt{n+1}}{-\Delta+2\Omega_n},
\end{equation}
where $\Omega_n=\sqrt{(\Delta/2)^2+g^2(n+1)}$ is the Rabi
frequency, and $\Delta=\omega_A-\omega_F$ is the detuning
parameter. In addition, corresponding to the eigensubspace
${\mathcal H}_{0}$, there exist a negative eigenvalue given by
\begin{equation}\label{EVneg}
E_{0}=-\frac{1}{2}\hbar \omega_A, \qquad
|\Phi_{0}\rangle=|g,0\rangle.
\end{equation}
Finally, taking into account equations
(\ref{Evalues}),(\ref{Evectors1}), (\ref{Evectors2}) and
(\ref{EVneg}), we arrive at the following equation for  the time
evolution operator $U(t)$
$$
U(t)={\rm e}^{i\omega_A t/2} |g,0\rangle \langle g,0|
+\sum_{n=0}^\infty\left({\rm e}^{-iE_{+}^{(n)}t/\hbar}
|\Phi_{+}^{(n)}\rangle\langle \Phi_{+}^{(n)}|\right.
$$
\begin{equation}\label{Ut}
\hspace{30mm}\left.+{\rm e}^{-iE_{-}^{(n)}t/\hbar}
|\Phi_{-}^{(n)}\rangle\langle \Phi_{-}^{(n)}|\right).
\end{equation}

\section{The time evolution of the system}
In order to study the entanglement properties of the JCM, let us
suppose that, initially at $t=0$, the system is found in the
product state
\begin{equation}
\rho(0)=\rho^A(0)\otimes \rho^F(0),
\end{equation}
such that $\rho^A(0)$, the initial state of the atom, is a general mixed
state with the diagonal representation
\begin{equation}
\rho^A(0)= \cos^2{(\frac{\varrho}{2})}|g\rangle\langle g | + \sin^2
{(\frac{\varrho}{2})}| e \rangle \langle e |,
\end{equation}
and $\rho^F(0)$, the initial state of the field,  is a general pure state
\begin{equation}
\rho^F(0)=|\eta\rangle\langle\eta|, \qquad|\eta\rangle =\sum_{n
=0}^\infty{b_n }|n \rangle,
\end{equation}
where the coefficients $b_n=\langle n|\eta\rangle$ are such that
the state is normalized, i.e.  $\sum_{n=0}^{\infty}|b_n|^2=1$. At
the end, we will fix the coefficients $b_n$ for the special case
that the initial state of the field is a coherent state
$|\alpha\rangle$.
 Accordingly, following the method given in \cite{furuichi3},
the final state of the system can be obtained as
$$\hspace{-11mm}
\rho(t)=  U(t)\rho(0)U^{\dag}(t)= \cos^2{(\frac{\varrho
}{2})}|\Psi_g(t)\rangle\langle \Psi_g(t)|
$$
\begin{equation}\label{Rhot}\hspace{36mm}
 +\sin^2{(\frac{\varrho
}{2})}|\Psi_e(t)\rangle\langle \Psi_e(t)|,
\end{equation}
where $|\Psi_g(t)\rangle$ and $|\Psi_e(t)\rangle$ are defined  by
\begin{equation}\label{psig}
|\Psi_g(t)\rangle=U(t)|g,\eta\rangle
=|g\rangle\otimes|\chi_1(t)\rangle+|e\rangle\otimes|\chi_2(t)\rangle,
\end{equation}
\begin{equation}\label{psie}
|\Psi_e(t)\rangle=U(t)|e,\eta\rangle
=|g\rangle\otimes|\chi_3(t)\rangle+|e\rangle\otimes|\chi_4(t)\rangle,
\end{equation}
where the unnormalized vectors $|\chi_\alpha(t)\rangle,\; (\alpha
= 1,\cdots,4)$ are given by
$$\hspace{-39mm}
|\chi_1(t)\rangle=\sum_{n = 0}^\infty b_{n}{\rm
e}^{-i\omega_F(n-1/2)t}
$$
\begin{equation}\label{chi1}\hspace{13mm}
\times\left(\cos{\Omega_{n-1}
t}-i\cos{2\vartheta_{n-1}}\sin{\Omega_{n-1} t}\right)|n\rangle,
\end{equation}
\begin{equation}\label{chi2}\hspace{-2mm}
|\chi_2(t)\rangle =-i\sum_{n=0}^\infty b_{n+1} {\rm
e}^{-i\omega_F(n+1/2)t}\sin{2\vartheta_n}\sin{\Omega_{n}t}
|n\rangle,
\end{equation}
\begin{equation}\label{chi3}
|\chi_3(t)\rangle =-i\sum_{n=0}^\infty b_n{\rm e}^{-i
\omega_F(n+1/2)t}\sin{2\vartheta_n}\sin{\Omega_n t}|n+1 \rangle,
\end{equation}
$$\hspace{-39mm}
|\chi_4(t)\rangle =\sum_{n=0}^\infty b_{n} {\rm
e}^{-i\omega_F(n+1/2)t}
$$
\begin{equation}\label{chi4}\hspace{-10mm}
\times\left(\cos{\Omega_n t
}+i\cos{2\vartheta_n}\sin{\Omega_{n}t}\right)|n\rangle.
\end{equation}
Alternatively, given an orthonormal basis $\{|e_1\rangle\equiv
|e\rangle,|e_2\rangle\equiv|g\rangle\}\in {\mathbb C}^2$ for the
atomic Hilbert space, we can represent the density matrix
$\rho(t)$  as
\begin{equation}\label{Rhot2}
\rho(t)=\left(\begin{array}{c|c} A(t) & C(t) \\ \hline
C^{\dag}(t) & B(t)
\end{array}\right),
\end{equation}
where $A(t)$, $B(t)$ and $C(t)$, operators acting on the field
Hilbert space, are defined by
\begin{equation}\label{At}
A(t)=\cos^2{(\frac{\varrho}{2})}|\chi_2(t)\rangle\langle
\chi_2(t)|+ \sin^2{(\frac{\varrho}{2})}|\chi_4(t)\rangle\langle
\chi_4(t)|,
\end{equation}
\begin{equation}\label{Bt}
B(t)=\cos^2{(\frac{\varrho}{2})}|\chi_1(t)\rangle\langle
\chi_1(t)|+ \sin^2{(\frac{\varrho}{2})}|\chi_3(t)\rangle\langle
\chi_3(t)|,
\end{equation}
\begin{equation}\label{Ct}
C(t)=\cos^2{(\frac{\varrho}{2})}|\chi_2(t)\rangle\langle
\chi_1(t)|+ \sin^2{(\frac{\varrho}{2})}|\chi_4(t)\rangle\langle
\chi_3(t)|.
\end{equation}
Therefore, in the basis $\{|e_1\rangle,|e_2\rangle\}$, the atomic
density matrix has the following matrix elements
$$\hspace{-22mm}\vspace{-3mm}
 \rho^{A}_{11}=\sum_{n=0}^{\infty}
\left[\cos^2(\frac{\varrho}{2})|b_{n+1}|^2\sin^2{2\vartheta_{n}}\sin^2{\Omega_{n}t}\right.
$$
\begin{equation} \hspace{5mm}\left.
+\sin^2(\frac{\varrho}{2})|b_{n}|^2\left(\cos^2{\Omega_{n}t}
+\cos^2{2\vartheta_{n}}\sin^2{\Omega_{n}t}\right)\right],
\end{equation}
$$\hspace{-10mm}
 \rho^{A}_{12}=-i{\textmd
e}^{-i\omega_Ft}\sum_{n=0}^{\infty}
\left[\cos^2(\frac{\varrho}{2})b_{n+1}b_n^\ast\sin{2\vartheta_{n}}\sin{\Omega_{n}t}\right.
$$
$$\left.
\times\left(\cos{\Omega_{n-1}t}+i\cos{2\vartheta_{n-1}}\sin{\Omega_{n-1}t}\right)\right.
$$
$$
-\left.\sin^2(\frac{\varrho}{2})b_{n}b_{n-1}^\ast\sin{2\vartheta_{n-1}}\sin{\Omega_{n-1}t}\right.
$$
\begin{equation}
 \left.
\times\left(\cos{\Omega_{n}t}+i\cos{2\vartheta_{n}}\sin{\Omega_{n}t}\right)\right],
\end{equation}
and $\rho^A_{21}=(\rho^A_{12})^\ast$, $\rho^A_{22}=1-\rho^A_{11}$.
The field density operator also takes the following form
$$\hspace{-55mm}
\rho^F(t)=A(t)+B(t)
$$
$$\hspace{-3mm}
=\cos^2{(\frac{\varrho}{2})}\left(|\chi_1(t)\rangle\langle
\chi_1(t)|+|\chi_2(t)\rangle\langle \chi_2(t)|\right) \\
$$
\begin{equation}\hspace{10mm}
+\sin^2{(\frac{\varrho}{2})}\left(|\chi_3(t)\rangle\langle
\chi_3(t)|+|\chi_4(t)\rangle\langle \chi_4(t)|\right),
\end{equation}

\section{Negativity}
In the subsequent sections our goal is to quantify the
entanglement of the final state (\ref{Rhot2}), using the concept
of the negativity defined by \cite{vidal}
\begin{equation}\label{Neg}
{\mathcal N}(\rho)\equiv \frac{\parallel
\rho^{T_1}\parallel_1-1}{2},
\end{equation}
where $\rho^{T_1}$ is the matrix obtained by partially transposing
the density matrix $\rho$ with respect to the first system, and
$\parallel \rho^{T_1}\parallel_1$ is the trace class norm of the
operator $\rho^{T_1}$. The trace class norm of any trace class
operator $A$ is defined by $\parallel
A\parallel_1=\rm{Tr}\sqrt{A^\dag A}$ \cite{conway}, which reduces
to the sum of the absolute value of the eigenvalues of $A$, when
$A$ is Hermitian. Therefore
\begin{equation}
\parallel\rho^{T_1}\parallel_1=\sum_{i}|\mu_i|
=\sum_{i}\mu_i-2\sum_{i}\mu_i^{neg}=1-2\sum_{i}\mu_i^{neg},
\end{equation}
where $\mu_i$ and $\mu_i^{neg}$ are, respectively,  the
eigenvalues and the negative eigenvalues of $\rho^{T_1}$. In the
last step, we used also the fact that
$\rm{Tr}\rho^{T_1}=\rm{Tr}\rho=1$.

The partial transposition of $\rho(t)$ with respect to the atom
in the basis $\{|e_1\rangle,|e_2\rangle\}\in {\mathbb C}^2$ is
defined by
\begin{equation}
\rho^{T_1}(t)=\sum_{i,j=1}^{2}\langle e_i|\rho(t)|e_j \rangle
|e_j\rangle \langle e_i |,
\end{equation}
where, using the representation (\ref{Rhot2}), can be written in
matrix form as
\begin{equation}\label{RhoT1}
\rho^{T_1}(t)=\left(\begin{array}{c|c} A(t) & C^{\dag}(t) \\
\hline C(t) & B(t)
\end{array}\right).
\end{equation}
Now in order to calculate the negativity we have to obtain the
eigenvalues of $\rho^{T_1}(t)$. If it was true that the set
$\{|\chi_\alpha(t)\rangle\}_{\alpha=1}^{4}$ is linearly
independent, then one can easily construct an orthonormal basis
set $\{|\phi_\alpha(t)\rangle\}_{\alpha=1}^{4}$ from them and
calculate the eigenvalues of $\rho^{T_1}(t)$ as a matrix in the
orthonormal basis $\{|e_i\rangle\otimes |\phi_\alpha\rangle\}$.
But this is not the case for all times, for instance at $t=0$ we
have only one independent vector
$|\chi_1(0)\rangle=|\chi_4(0)\rangle=|\eta\rangle$ with
$|\chi_2(0)\rangle=|\chi_3(0)\rangle=0$. This, as we know from the
initial conditions, implies that $\rho(0)$ acts  on ${\mathbb
C}^2\otimes {\mathbb C}^1$. However, due to the time evolution of
the state and generation of entanglement, it is evident from
numerical calculations that for $t> 0$ the dependency between
vectors is destroyed and,
 except for some points, we have four linearly independent vectors.
It follows that  at any time $t$ we can always find an
$N$-dimensional subspace in the field Hilbert space ${\mathcal
H}^F$ such that $1\le N\le 4$ denotes the number of independent
vectors in the set $\{|\chi_\alpha(t)\rangle\}_{\alpha=1}^{4}$.
Therefore the field density operator acts instantaneously on this
$N$-dimensional subspace and $\rho(t)$ has a support on ${\mathbb
C}^2\otimes {\mathbb C}^N$, consequently, the number of nonzero
eigenvalues of $\rho^{T_1}(t)$ are at most eight.  On the other
hand we can calculate the eigenvalues of $\rho^{T_1}(t)$ by
expanding the field operators of equation (\ref{RhoT1}) in the
number states $\{|n\rangle\}_{n=0}^{n=\infty}$ as
$$\hspace{-10mm}
A_{mn}= \exp{\left(-i\omega_{F}(m-n)t\right)}
\left[\cos^2{(\frac{\varrho}{2})}b_{m+1}b^\ast_{n+1}\right.
$$
$$
\left.\times\sin{2\vartheta_{m}}\sin{2\vartheta_{n}}\sin{\Omega_{m}t}\sin{\Omega_{n}t}\right.
$$
$$
+  \left.\sin^2{(\frac{\varrho}{2})}b_{m}b^\ast_{n}
\left(\cos{\Omega_{m}t}+i\cos{2\vartheta_{m}}\sin{\Omega_{m}t}\right)\right.
$$
\begin{equation}\label{Amn}
\times\left.\left(\cos{\Omega_{n}t}-i\cos{2\vartheta_{n}}\sin{\Omega_{n}t}\right)\right],
\end{equation}

$$\hspace{-10mm}
B_{mn}= \exp{\left(-i\omega_{F}(m-n)t\right)}
\left[\sin^2{(\frac{\varrho}{2})}b_{m-1}b^\ast_{n-1}\right.
$$
$$
\left.\times
\sin{2\vartheta_{m-1}}\sin{2\vartheta_{n-1}}\sin{\Omega_{m-1}t}\sin{\Omega_{n-1}t}\right.
$$
$$
+ \cos^2{(\frac{\varrho}{2})}b_{m}b^\ast_{n}
\left.\left(\cos{\Omega_{m-1}t}-i\cos{2\vartheta_{m-1}}\sin{\Omega_{m-1}t}\right)\right.
$$
\begin{equation}\label{Bmn}
\times\left.\left(\cos{\Omega_{n-1}t}+i\cos{2\vartheta_{n-1}}\sin{\Omega_{n-1}t}\right)\right],
\end{equation}

$$
C_{mn}=  -i\exp{\left(-i\omega_{F}(m-n+1)t\right)}
\times\left[\cos^2{(\frac{\varrho}{2})}b_{m+1}b^\ast_{n}\right.
$$
$$
\left.\times\sin{2\vartheta_{m}}\sin{\Omega_{m}t}
\left(\cos{\Omega_{n-1}t}+i\cos{2\vartheta_{n-1}}\sin{\Omega_{n-1}t}\right)\right.
$$
$$
-\sin^2{(\frac{\varrho}{2})}b_{m}b^\ast_{n-1}
\sin{2\vartheta_{n-1}}\sin{\Omega_{n-1}t}
$$
\begin{equation}\label{Cmn}
\times\left.\left(\cos{\Omega_{m}t}+i\cos{2\vartheta_{m}}\sin{\Omega_{m}t}\right)
\right],
\end{equation}
and $(C^{\dag})_{mn}=C^{\ast}_{nm}$.  Although this
representation gives an infinite-dimensional  matrix for
$\rho^{T_1}(t)$ which should be solved for its eigenvalues, but
numerical calculations show that for large values of $m$ and $n$
the above matrix elements rapidly tends to zero. For instance when
the  field density is $|\alpha|=\sqrt{5}$ we find that $|b_n|^2$
is less than $10^{-10}$ for $n>25$. Therefore the calculation of
the eigenvalues of $\rho^{T_1}(t)$ can be easily handled in the
truncated finite-dimensional  matrix representation of
$\rho^{T_1}(t)$ for some $n$. In all cases, however, we have only
at most eight nonzero eigenvalues for $\rho^{T_1}(t)$.

Horodecki et al  have shown that if the rank of a density matrix
is equal to two, then the positive partial transpose is a
necessary and sufficient condition for separability \cite{horo1}.
As we mentioned above (see equation (\ref{Rhot})) the final state
of the JCM is a rank two density matrix supported at most on
${\mathbb C}^2\otimes{\mathbb C}^4$. This means that the PPT
condition is a necessary and sufficient condition for
separability of $\rho(t)$. Consequently, the negativity of the
partial transpose fully captures the entanglement of $\rho(t)$,
that is, ${\mathcal N}(\rho(t))=0$ iff $\rho(t)$ is separable.

\section{Numerical results and discussion}
\begin{figure}[h]\label{fig1}
\centerline{\includegraphics[height=11cm]{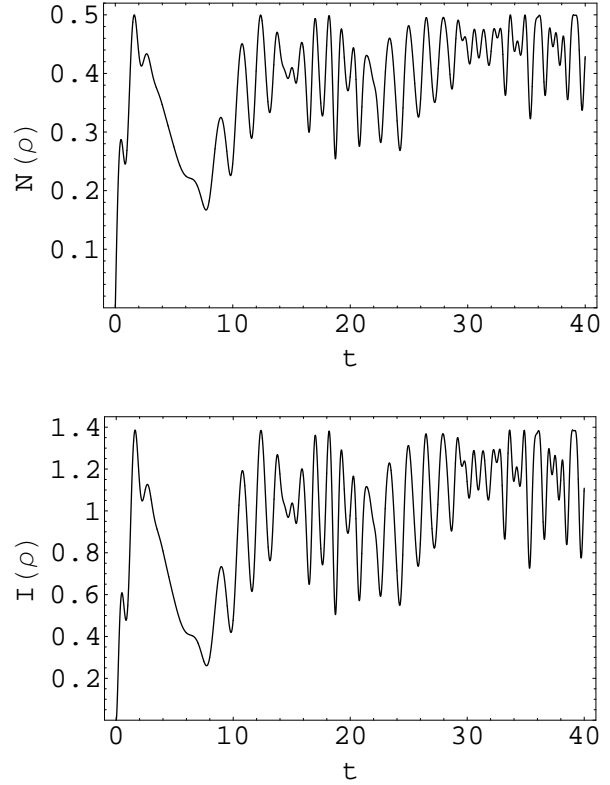}}
\caption{The time evolution of the negativity and the mutual
entropy for $\alpha=\sqrt{5}$, $g=1$, $\omega_A=1$, $\Delta=0$
and $\cos^2{(\frac{\varrho}{2})}=0$.}
\end{figure}

\begin{figure}[h]\label{fig2}
\centerline{\includegraphics[height=11cm]{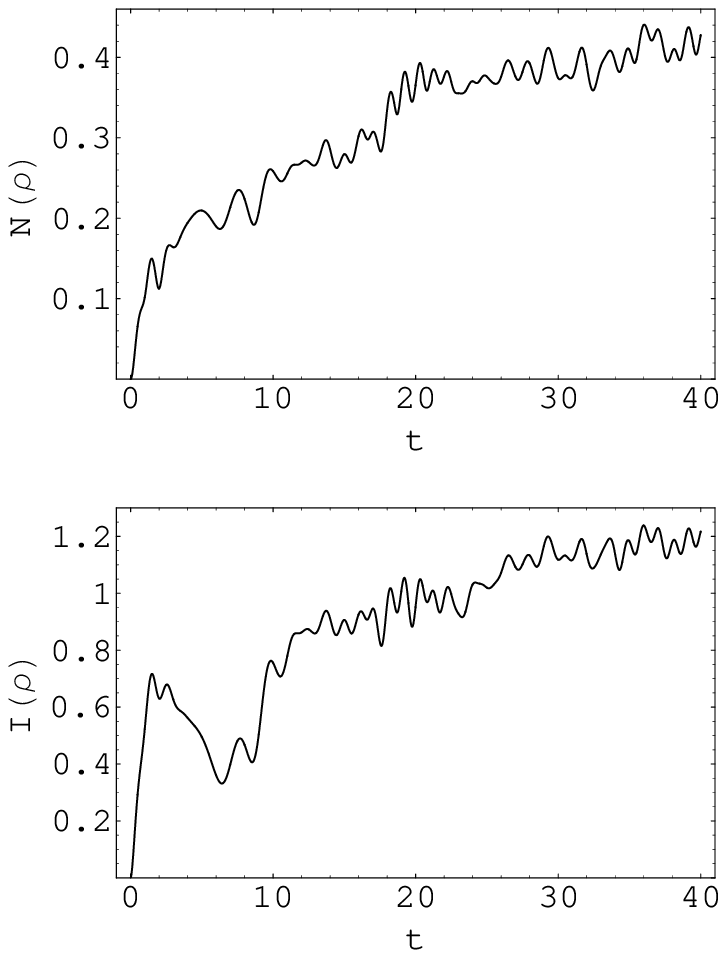}}
\caption{The time evolution of the negativity and the mutual
entropy for $\alpha=\sqrt{5}$, $g=1$, $\omega_A=1$, $\Delta=0$
and $\cos^2{(\frac{\varrho}{2})}=1/2$.}
\end{figure}

\begin{figure}[h]\label{fig3}
\centerline{\includegraphics[height=11cm]{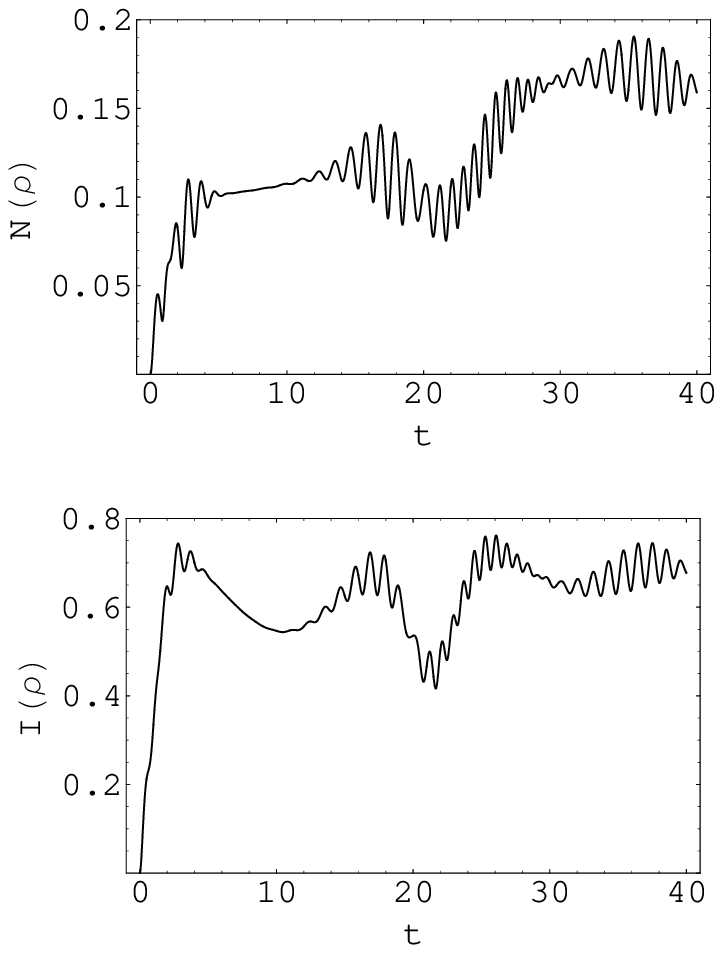}}
\caption{The time evolution of the negativity and the mutual
entropy for $\alpha=\sqrt{5}$, $g=1$, $\omega_A=1$, $\Delta=5$
and $\cos^2{(\frac{\varrho}{2})}=1/2$.}
\end{figure}

\begin{figure}[h]\label{fig4}
\centerline{\includegraphics[height=11cm]{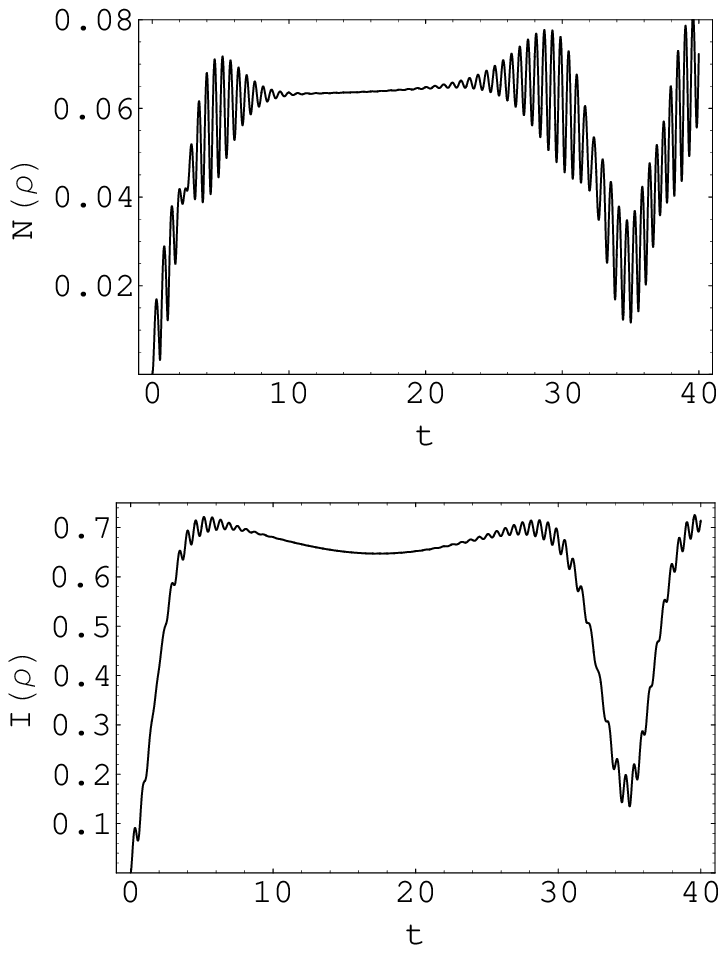}}
\caption{The time evolution of the negativity and the mutual
entropy for $\alpha=\sqrt{5}$, $g=1$, $\omega_A=1$, $\Delta=10$
and $\cos^2{(\frac{\varrho}{2})}=1/2$.}
\end{figure}

\begin{figure}[h
]\label{fig5}
\centerline{\includegraphics[height=4.5cm]{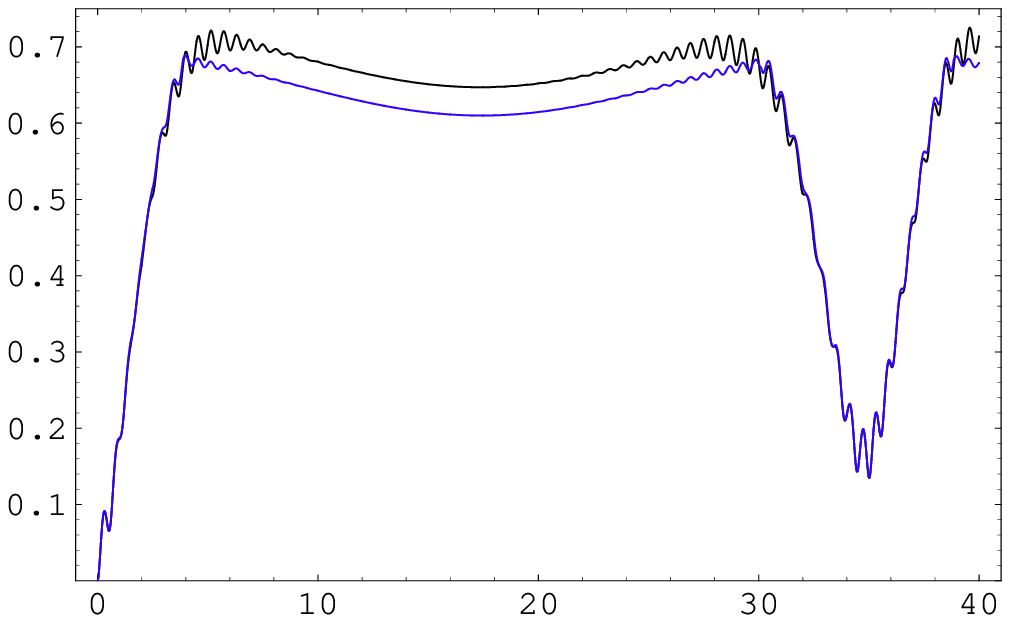}}
\caption{The time evolution of the mutual entropy (upper curve)
and the classical upper bound (lower curve)  for
$\alpha=\sqrt{5}$, $g=1$, $\omega_A=1$, $\Delta=10$ and
$\cos^2{(\frac{\varrho}{2})}=1/2$.}
\end{figure}

In what follows we are going to calculate the negativity by
considering special cases for the state of the field. Although
${\rm rank}(\rho^{T_1}(0))={\rm rank}(\rho(0))=2$, but as
mentioned in the last section, due to the time evolution of the
state and generation of entanglement, the rank of $\rho^{T_1}(t)$
is no longer equal to 2 and we may have in general $2\le{\rm
rank}(\rho^{T_1}(t))\le 8$. Therefore it is not possible to
calculate the negativity analytically and numerical calculation is
required. We now suppose the initial state of the field is a
coherent state $|\alpha\rangle=\sum_{n=0}^{\infty}b_n|n\rangle$
with $b_n=e^{-|\alpha|^2/2}\frac{\alpha^n}{\sqrt{n!}}$
\cite{scully}. This system when the atom is also in a pure state
is studied by Phoenix et al \cite{phoenix}, where they have used
the von Neumann entropy of the reduced density matrix in order to
describe the entanglement properties. Also by using the quantum
mutual entropy, Furuichi et al \cite{furuichi3} have studied this
system with the atom in a mixed state. Our numerical results of
the negativity and the mutual entropy are shown in figures
(1)-(5). Figure (1) shows the negativity and the mutual entropy
when the atom is initially in the excited state, i.e.
$\cos^2{(\frac{\varrho}{2})}=0$, and without the detuning . In
this particular case the state of the whole bipartite system is
pure and therefore the von Neumann entropy of atom (or field)
gives the entanglement of the state, and remarkably, the mutual
entropy gives twice the reduced von Neumann entropy, i.e.
$I(\rho(t))=2S(\rho(t))$. Surprisingly, up to a difference
between the amplitude of oscillation, there exist a complete
agreement between ${\mathcal N}(\rho)$ and $I(\rho)$ for this
case. It is noteworthy that since in this case $\rho(t)$ is
supported on $2\otimes 2$ space, therefore the entanglement of
formation can be calculated exactly using the Wootters's
concurrence formula \cite{woot}. Let us now consider the case
that $\cos^2{(\frac{\varrho}{2})}\neq 0,1$, i.e. the initial
state of the atom is not a pure state and, instead, it is a
statistical mixture of the ground and the excited states. Figure
(2) shows the time development of the negativity and the mutual
entropy for $\cos^2{(\frac{\varrho}{2})}=1/2$, $\Delta=0$ and
field intensity $|\alpha|^2=5$. Comparing ${\mathcal N}(\rho)$
and $I(\rho)$ in this figure one finds that the agreement between
two measures diminish as long as we deal with a mixed-state
density matrix. It follows that the main difference between the
two measures occurs during the collapse region $t\cong 3-7$ where
the two measures do not show the same ordering for entanglement.
Figures (3) and (4) are plotted the same as in figure (2) but,
respectively, with the non zero detuning $\Delta=5$ and
$\Delta=10$. Again it is apparent that the two measures do not
show the same behavior.

Comparing figures (2), (3) and (4), it is clear that by increasing
the detuning parameter $\Delta$, the negativity decreases faster
than the mutual entropy, therefore the relative difference
between their amplitudes of  oscillation increases. Numerical
calculations (see figure (5)) show that for parameters of figure
(4), i.e. for $\Delta=10$, the mutual entropy $I(\rho)$ is
approximately equal to the classical upper bound
$\min{[S(\rho^A),S(\rho^F)]}$, i.e. the bound that becomes
saturated when the two  subsystems $A$ and $F$ are classically
maximally correlated. Since only the range between classical and
quantum upper bounds corresponds to pure quantum entanglement
\cite{cerf}, i.e. $\min{[S(\rho^A),S(\rho^F)]}\le I(\rho)\le
2\min{[S(\rho^A),S(\rho^F)]}$, therefore it seems that for large
values of the detuning, the mutual entropy evaluation of the
system shows that the  system is more classically correlated than
quantum correlated.

 \section{Conclusion}
 In this paper we have used the negativity in order to quantify
the  entanglement of the JCM. We have supposed that the initial
state of the atom is
 a general mixed state, and the field is initially in a coherent
 state.
 It is shown that the state operator
$\rho(t)$ acts on ${\mathbb C}^2\otimes {\mathbb C}^N$ where $1\le
N\le 4$.
 In this case the negativity fully captures the entanglement
 properties of the system. It is investigated, with the help of numerical
 calculations,  that  when the system state is pure, the negativity shows
the  same functionality, up to a difference in the amplitude of
oscillation,
 with the reduced von Neumann entropy (and also mutual entropy)
 which is widely accepted as  the entanglement measure of pure states.
  On the other hand for mixed states
 one finds that the
 agreement between the negativity and the mutual entropy,
 which have been used in the most of the previous studies on the entanglement of the JCM,
 is diminished. The effect of the detuning is also
 examined, and it is shown that
 the quantum entanglement evaluated by the negativity has a small value for
the large detuning. We have also seen that for large values of the
detuning, the mutual entropy evaluation of the system shows that
the system is more classically correlated than quantum correlated.
Since for any rank $N$  bipartite system supported on
 ${\mathcal C}^2\otimes {\mathcal C}^N$, the positivity of the partial
 transpose is necessary and sufficient condition for separability \cite{kraus}, therefore
 the negativity may be used also to study the entanglement properties of the JCM
  in  the cases that the initial state of the
 field is not pure and instead it is, for instance, a non coherent mixture
 of the two coherent states $|\alpha\rangle$ and $|-\alpha\rangle$.

At the end, it should be mentioned that various physical systems
have been suggested to realize the generation of entangled states
of atom and radiation, where the micromaser is one of the
experimental and theoretical  realization of such systems
\cite{mesched,filip,rempe,walther}. Another fundamental system to
realize the JCM and generation of entangled states is a trapped
two-level ion interacting with a laser beam, where the laser beam
couples the quantized internal states of the ion to the quantized
center-of-mass motional states of the ion
\cite{wine,died,monroe1,cirac,monroe2,sackett,rowe,sharma}.


\begin{thebibliography}{99}
\bibitem{ben1}{ C. H. Bennett, and S. J. Wiesner,}
{\em Phys. Rev. Lett. {\bf 69}, 2881 (1992).}
\bibitem{ben2}{ C. H. Bennett, G. Brassard,
C. Cr\'{e}peau, R. jozsa, A Peres and W. K. Wootters,}
{\em Phys.
Rev. Lett. {\bf 70}, 1895 (1993).}
\bibitem{ben3}{ C. H. Bennett, D. P. DiVincenzo, J. A. Smolin and W.K.
Wootters,} {\em Phys. Rev. A {\bf 54}, 3824 (1996).}
\bibitem{werner}{ R. F. Werner, }{\em Phys. Rev. A {\bf 40} 4277 (1989).}
\bibitem{peres}{ A. Peres, }{\em Phys. Rev. Lett. {\bf 77} 1413 (1996).}
\bibitem{horo0}{ M. Horodecki, P. Horodecki and R. Horodecki, }
{\em Phys. Lett. A  {\bf 223} 1 (1996).}
\bibitem{woot}{ W. K. Wootters, }
{\em Phys. Rev. Lett. {\bf 80} 2245 (1998).}
\bibitem{ved1}{ V. Vedral, M. B. Plenio, M. A. Rippin and P. L. Knight,}
 {\em Phys. Rev. Lett. {\bf 78}, 2275 (1995).}
\bibitem{ved2}{ V. Vedral and M. B. Plenio,}
 {\em Phys. Rev. A {\bf 57}, 1619 (1998).}
\bibitem{zycz1}{ K. Zyczkowski, P. Horodecki, A. Sanpera and M. Lewenstein, }
{\em Phys. Rev. A {\bf 58} 883 (1998).}
\bibitem{zycz2}{ K. Zyczkowski, }
{\em Phys. Rev. A {\bf 60} 3496 (1999).}
\bibitem{vidal}{ G. Vidal and R. F. Werner, }
{\em Phys. Rev. A {\bf 65} 032314 (2002).}
\bibitem{JC}{ E. T. Jaynes and F. W. Cummings, }
{\em Proc. IEEE {\bf 51} 89 (1963).}
\bibitem{phoenix}{ S. J. D. Phoenix and P. L. Knight, }
{\em Phys. Rev. A {\bf 44} 6023 (1991).}
\bibitem{furuichi4}{ S. Furuichi and M. Abdel-Aty, }
{\em J. Phys. A: Math. Gen. {\bf 34} 6851 (2001).}
\bibitem{furuichi3}{ S. Furuichi and S. Nakamura, }
{\em J. Phys. A: Math. Gen. {\bf 35} 5445 (2002).}
\bibitem{scheel}{S. Scheel, J. Eisert, P. L. Knight and M. B. Plenio,}
{\em J. Mod. Opt. {\bf 50} 881 (2003).}
\bibitem{raja}{A. K. Rajagopal, K. L. Jensen and F. W. Cummings, }
{\em Phys. Lett. A {\bf 259} 285 (1999).}
\bibitem{khem}{S. Khemmani, V. Sa-Yakanit and W. T. Strunz,}
{\em Phys. Lett. A {\bf 341} 87 (2005).}
\bibitem{fang}{ Mao-Fa Fang and Peng Zhou, }
{\em Physica A {\bf 234} 571 (1996).}
\bibitem{abdel1}{ M. Abdel-Aty, S. Furuichi and A-S F. Obada }
{\em J. Opt. B: Quantum Semiclass. {\bf 4} 37 (2002).}
\bibitem{abdel2}{ M. Abdel-Aty, }
{\em J. Math. Phys. {\bf 44} 1457 (2003).}
\bibitem{abdel4}{M. Abdel-Aty,}
{\em Physica A {\bf 313} 471 (2002).}
\bibitem{li}{ S.-B. Li and J.-B. Xu, }
{\em Phys. Lett. A {\bf 337} 321 (2005).}
\bibitem{abdalla}{ M. S. Abdalla, M. Abdel-Aty, and A.-S. F. Obada }
{\em Physica A {\bf 326} 203 (2003).}
\bibitem{zhou}{ L. Zhou, H. S. Song and C. Li, }
{\em J. Opt. B: Quantum Semiclass. {\bf 4} 425 (2002).}
\bibitem{abdel3}{ M. Abdel-Aty and A.-S. F. Obada, }
{\em J. Math. Phys. {\bf 45} 4271 (2004).}
\bibitem{wang}{C.-Z. Wang, C.-X. Li and G.-C. Guo,}
{\em Euro. Phys. J. D {\bf 37} 267 (2006).}
\bibitem{obada}{ A.-S. F. Obada and M. Abdel-Aty,}
{\em Euro. Phys. J. D {\bf 27} 277 (2003).}
\bibitem{zhou2}{ L. Zhou, H. S. Song, Y. X. Luo and C. Li, }
{\em Phys. Lett. A {\bf 284} 156 (2001).}
\bibitem{zhou3}{ L. Zhou, H. S. Song and Y. X. Luo, }
{\em J. Opt. B: Quantum Semiclass. {\bf 4} 103 (2002).}
\bibitem{zidan}{N. A. Zidan, M. Abdel-Aty and and A.-S. F. Obada, }
{\em Chaos, Solitons and Fractals, {\bf 13} 1421 (2002).}
\bibitem{ren}{R. W. Rendell and A. K. Rajagopal,}
{\em Phys. Rev. A {\bf 67} 062110 (2003).}
\bibitem{bouk}{ E. Boukobza and D. J. Tannor, }
{\em Phys. Rev. A {\bf 71} 063821 (2005).}
\bibitem{eisert}{ J. Eisert and M. Plenio, }
{\em J. Mod. Opt.  {\bf 46} 145 (1999).}
\bibitem{wei}{T-C Wei, K. Nemoto, P. M. Goldbart,P. G. Kwiat, W. J. Munro and F. Verstraete,}
{\em Phys. Rev. A {\bf 67} 022110 (2003).}
\bibitem{miran1}{A. Miranowicz and A. Grudka, }
{\em Phys. Rev. A {\bf 70} 032326 (2004).}
\bibitem{barnett}{ S. M. Barnett and S. J. D. Phoenix, }
{\em Phys. Rev. A {\bf 40} 2404 (1989).}
\bibitem{lindblad}{ G. Lindblad, }
{\em Commun. Math. Phys. {\bf 33} 305 (1973).}
\bibitem{cerf}{N. J. Cerf and C. Adami, }
{\em Phys. Rev. Lett. {\bf 79} 5194 (1997).}
\bibitem{ved3}{ V. Vedral, }
{\em Rev. Mod. Phys. {\bf 74} 197 (2002).}
\bibitem{horo1}{ P. Horodecki, J. S. Smolin, B. M. Terhal and A. V. Thapliyal, }
{\em Theo. Comp. Sci. {\bf 292} 589 (2003).}
\bibitem{kraus}{ B. Kraus, J. I. Cirac, S. Karnas and M. Lewenstein, }
{\em Phys. Rev. A {\bf 61} 062302 (2000).}
\bibitem{conway}{\sc J. B. Conway,} { \em ``A Course in Operator
Theory''}, American Mathematical Society, (2000)
\bibitem{scully}{ M. O. Scully, M. S. Zubairy,}
{\em ``Quantum optics''}, Cambridge university press, (1997).
\bibitem{mesched}{D. Mesched, H. Walther and G. Muller, }
{\em Phys. Rev. Lett. {\bf 54} 551 (1985).}
\bibitem{filip}{P. Filipovicz, J. Javanaein and P. Meystre, }
{\em Phys. Rev. A {\bf 34} 3077 (1986).}
\bibitem{rempe}{G. Rempe, H. Walther and N. Klein, }
{\em Phys. Rev. Lett. {\bf 58} 353 (1987).}
\bibitem{walther}{H. Walther, }
{\em Physica Scripta {\bf T23} 165 (1988).}
\bibitem{wine}{D. J. Wineland, W. M. Itano, J. C. Berguist and R. G. Hulet,}
{\em Phys. Rev. A {\bf 36} 2220 (1987).}
\bibitem{died}{F. Diedrich, J. C. Berguist, W. M. Itano, D. J. Wineland,}
{\em Phys. Rev. Lett. {\bf 62} 403 (1989).}
\bibitem{monroe1}{C. Monroe, D. M. Meekhof, B. E. King, J. R. Jefferts,
 W. M. Itano, D. J. Wineland,}
{\em Phys. Rev. Lett. {\bf 75} 4011 (1995).}
\bibitem{cirac}{J. I. Cirac and p. zoller,}
{\em phys. Rev. Lett. {\bf 74} 4091 (1995).}
\bibitem{monroe2}{C. Monroe, D. M. Meekhof, B. E. King and D. J. Wineland,}
{\em Science  {\bf 212} 1131 (1996).}
\bibitem{sackett}{C. A. Sackett, D. Kielpinski, B. E. King, C. Langer, V. Meyer, C. J. Myatt, M. Rowe,
Q. A. Turchette, W. M. Itano, D. J. Wineland and I.C. Monroe,}
{\em Nature {\bf 404} 256 (2000).}
\bibitem{rowe}{M. Rowe, D. Kielpinski, V. Meyer, C. A. Sackett,
 W. M. Itano, I.C. Monroe and D. J. Wineland,} {\em Nature {\bf
409} 791 (2001).}
\bibitem{sharma}{S. S. Sharma, E. De Almeida and N. K. Sharma,}
{\em J. Phys. B: At. Mol. Opt. Phys. {\bf 39} 695 (2006).}

\end{thebibliography}
\end{document}